\documentclass[aps,prd,reprint,amsmath,amssymb,nofootinbib]{revtex4-2}
\usepackage{graphicx}
\ifdefined\pdfinfoomitdate\pdfinfoomitdate=1\fi
\usepackage{bm}
\usepackage[english]{babel}
\usepackage[hidelinks]{hyperref}
\newif\ifreplicationattached
\IfFileExists{Guo_Herts_PRD_v4_data.zip}{%
  \IfFileExists{embedfile.sty}{\usepackage{embedfile}\replicationattachedtrue}{}%
}{}
\AtBeginDocument{%
  \ifreplicationattached
    \embedfile[stringmethod=escape,mimetype=application/zip,desc={Synthetic lensing data and reproducible analysis}]{Guo_Herts_PRD_v4_data.zip}%
  \fi
}
\hypersetup{pdftitle={Testing single-screen lensing with chromatic astrometry and delay drifts},pdfauthor={Jiabao Guo},pdfsubject={Thin-screen lensing and dispersive radio propagation},pdfkeywords={gravitational lensing, plasma lensing, astrometry, group delay}}
\newcommand{\thv}{\bm{\theta}}
\newcommand{\bv}{\bm{\beta}}
\newcommand{\gv}{\bm g}
\newcommand{\hv}{\bm h}
\newcommand{\Bmat}{\mathsf B}
\newcommand{\dd}{\mathrm d}
\newcommand{\tr}{^{\mathsf T}}
\newcommand{\tst}{t_\star}

\newcommand{\prj}{\mathsf{\Pi}}
\graphicspath{{figures/}{./}}
\begin{document}
\title{Testing single-screen lensing with chromatic astrometry and delay drifts}
\author{Jiabao Guo}
\email[Corresponding author: ]{jg26aai@herts.ac.uk}
\affiliation{University of Hertfordshire, Hatfield, United Kingdom}

\begin{abstract}
Can measured radio image positions predict how the delay between the images changes as a source moves? For a single static thin screen, we show that the image separations averaged along the source path at the two band edges determine this change exactly. The relation allows arbitrary smooth dispersion. An unknown source displacement introduces two common parameters that can be fitted across image pairs and bands. With two screens, timing probes where the rays cross the last screen. For a quadratic foreground plasma we derive the observable projection of its curvature and give a condition for its recovery. At a fixed source position we construct different gravitational and plasma potentials with identical image tracks and group delays. A four-image numerical experiment quantifies the angular precision needed to detect a second screen, and a three-image plasma lens yields a positive drift for one image pair in every tested band despite a negative band-center prediction.
\end{abstract}
\maketitle

\section{Introduction}

A repeating radio source allows the delay between two resolved images to be measured at several epochs. Matching the same emission event in the two images gives their differential group delay. Matching a later event gives the change in that delay as the source and lens move relative to one another. In vacuum the contribution of source motion is proportional to image separation, even when the lens potential is unknown \cite{DaiLu,Zitrin,Wucknitz}. In a plasma the separation changes across the observing band. Using the angle at the band center can then introduce a systematic error even for a stationary single lens.

Several lines of work motivate this question. A compact mass can create two echoes of a fast radio burst (FRB) \cite{Munoz}. The changing delays of strongly lensed repeaters have been studied as probes of source motion and cosmological evolution \cite{DaiLu,Zitrin,Wucknitz}. At radio frequencies, intrinsic core shift can give an active galactic nucleus different phase and group-delay positions \cite{Lobanov,Porcas}. In pulsars the derivatives of a common interference phase give the pair delay and fringe rate \cite{Cordes}. Refractive plasma lenses can form multiple images of Galactic and extragalactic radio sources, even in the absence of a compact gravitating lens \cite{Clegg,PenKing,CordesFRB,Simard}. Dynamic spectra constrain a plasma lens along characteristics \cite{Tuntsov}. Observations of eclipsing and scintillating pulsars give examples of resolved emission structure and localized scattering paths \cite{Main,Brisken,Sprenger}.

Wagner and Er used positions and delays at nearby wavelengths to constrain distance and electron-column combinations \cite{Wagner}. Er and Mao and Bisnovatyi-Kogan and Tsupko calculated timing changes in model gravitational lenses with plasma \cite{ErMao,BKTtime}. Multi-plane Fermat theory describes stationary rays and their caustics \cite{Blandford,Petters,Schneider}, and double-plane plasma calculations predict positions, magnifications and pulse profiles \cite{ErDouble}. These results leave an observational question. If the mass map, electron column and source proper motion are all unknown, which combinations of the measured delays and angles must still agree?

Our results are as follows.
\begin{enumerate}
\item Equation~\eqref{eq:finite} expresses a finite change in band-averaged pair delay in terms of the path-mean angular separations at the band edges. Repeated bands and pairs permit a test after fitting a common displacement.
\item For a quadratic foreground plasma, Eq.~\eqref{eq:extended} identifies the curvature projection measured by one source trajectory. Equation~\eqref{eq:info} gives the part that survives fitting the motion. An explicit achromatic foreground passes the one-screen test exactly.
\item At a fixed source position, Eq.~\eqref{eq:ambiguity} constructs smooth changes to the gravitational and plasma potentials that preserve the measured image tracks and group delays over a finite band.
\item A four-image simulation quantifies the precision loss and the bias from fitting a spatially varying foreground with a constant curvature. A separate plasma-only example in Sec.~\ref{sec:pulsars} predicts the sign and magnitude of a delay change without a gravitational mass model.
\end{enumerate}

\section{A finite-band relation for a single screen}
\label{sec:single}
\begin{figure*}[t]
 \centering\includegraphics[width=.97\textwidth]{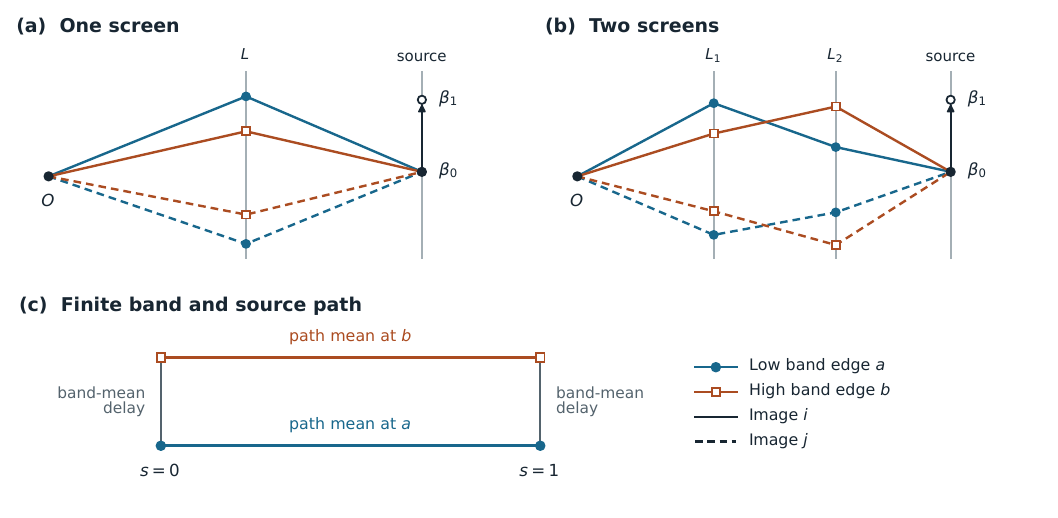}
 \caption{\label{fig:geometry}Geometry and measurements. (a) Two stationary images of a source at $\bv_0$ are shown at band edges $a$ and $b$. At a single screen the observed angles identify the screen crossings. The arrow follows the source to $\bv_1$. (b) With two screens the measured angles refer to plane 1, whereas the source derivative of the phase selects the angular coordinates on plane 2. Plane coordinates are transverse positions divided by the corresponding observer-to-plane distances. (c) Horizontal lines average image separations along the source path at each band edge. Vertical lines average group delays over frequency at each endpoint of the path.}
\end{figure*}

\subsection{Source position and frequency derivatives}

Let $\nu$ be the observer's ordinary frequency. In a fixed angular reference frame a single static thin screen has a phase in cycles $\nu F$, with
\begin{equation}
 F=\frac K2|\thv-\bv|^2+V(\thv,\nu),\quad
 K=\frac{(1+z_L)D_LD_S}{cD_{LS}} .
 \label{eq:screen}
\end{equation}
Angles are in radians and $F,V,K$ have time units. The distance factors in Eq.~\eqref{eq:screen} are angular-diameter distances. A term depending only on $\bv$ and $\nu$ cancels from our image differences. $V$ contains the gravitational screen $-K\psi$ and the material phase, expressed in observer-time units. It can depend nonlinearly on frequency. A scalar, real phase, a pointlike emitter, and a matched stationary branch are assumed. For two source positions the lens and observer are taken to be fixed, or the single screen has been reduced to this frame with a common relative trajectory. Cosmological changes in $K$ during the measurement are neglected.

For each image $i$, $\nabla_\theta F=0$ determines $\thv_i(\bv,\nu)$. Define $f_i=F(\thv_i,\bv,\nu)$ and $t_i=\partial_\nu(\nu f_i)$ at fixed $\bv$. An image difference is always $i$ minus $j$. Suppose the phase is smooth and the image Hessians remain invertible over the band and source scan. For a source component $\beta_l$, the total derivative along the stationary image is
\begin{align*}
 \partial_{\beta_l} f_i
 &= F_{,\beta_l}|_i
    +(\nabla_\theta F|_i)\cdot\partial_{\beta_l}\thv_i\\
 &=K(\beta_l-\theta_{i,l}).
\end{align*}
The second term vanishes by stationarity. Subtracting two images removes the source coordinate and yields
\begin{equation}
 \nabla_\beta f_i=K(\bv-\thv_i),\qquad
 \nabla_\beta\Delta f=-K\Delta\thv .
 \label{eq:envelope}
\end{equation}
To obtain the group delay, the derivative is taken at fixed source position. Since the phase is stationary in angle, $t_i=f_i+\nu V_{,\nu}|_i$. The second term still depends on the branch. Define the symmetric Hessian $H_i=K I+V_{,\theta\theta}|_i$ and let $\bm e_l$ be the $l$th basis vector. Differentiating $K(\thv_i-\bv)+\nabla_\theta V|_i=0$ gives
\begin{align*}
 H_i\,\partial_{\beta_l}\thv_i&=K\bm e_l,&
 H_i\,\partial_\nu\thv_i&=-\nabla_\theta V_{,\nu}|_i .
\end{align*}
Consequently the angular contribution to $\partial_{\beta_l}t_i$ is
\[
 \nu(\nabla_\theta V_{,\nu}|_i)\tr H_i^{-1}K\bm e_l
       =-K\nu\,\partial_\nu\theta_{i,l}.
\]
Adding $\partial_{\beta_l}f_i$ and taking the image difference gives
\begin{equation}
 \boxed{\ \nabla_\beta\Delta t
       =-K\,\partial_\nu\!\left(\nu\,\Delta\thv\right).\ }
 \label{eq:local}
\end{equation}
Here $\Delta\thv$ is the separation determined from image directions, as in phase-referenced astrometry. A broadband interferometric group-delay position may already contain the frequency derivative in Eq.~\eqref{eq:local} \cite{Porcas}. Substituting it for an independent phase position would count that effect twice. In vacuum Eq.~\eqref{eq:local} reduces to the transverse source-motion term in Ref.~\cite{Wucknitz}. If $\Delta\thv=\bm a+\bm b(\nu_\star/\nu)^2$, the combination on the right is $\bm a-\bm b(\nu_\star/\nu)^2$. A component proportional to $\nu^{-1}$ drops out. Both examples follow without solving a lens equation.

\subsection{Finite epochs, finite bands and rank}
\label{sec:finite}
Write the source trajectory as $\bv(s)=\bv_0+s\,\delta\bv$, $0\leq s\leq1$, where $s$ is a normalized epoch. Define the mean image separation along that trajectory and the change in the measured band delay by
\begin{align}
 \overline{\Delta\thv}(\nu)&=\int_0^1\Delta\thv(\bv(s),\nu)\,\dd s,\\
 d_{ab}&=\frac{1}{b-a}\int_a^b
      [\Delta t(\bv_1,\nu)-\Delta t(\bv_0,\nu)]\,\dd\nu ,
 \label{eq:ddef}
\end{align}
where $a,b$ are frequency edges with $0<a<b$. Define a vector for each band and pair,
\begin{equation}
 \bm B_{ab}=\frac{b\,\overline{\Delta\thv}(b)
                    -a\,\overline{\Delta\thv}(a)}{b-a}.
 \label{eq:Bdef}
\end{equation}
At each frequency the endpoint difference of $\Delta t$ is the integral of $\delta\bv\cdot\nabla_\beta\Delta t$ along the source trajectory. Substitution of Eq.~\eqref{eq:local} gives
\begin{align*}
 d_{ab}
  &=-\frac{K}{b-a}\int_0^1\dd s\,\delta\bv\cdot
       \int_a^b\dd\nu\,\partial_\nu(\nu\Delta\thv)\\
  &=-\frac{K}{b-a}\,\delta\bv\cdot
       \int_0^1[\,\nu\Delta\thv(\bv(s),\nu)\,]_a^b\,\dd s .
\end{align*}
Both band edges retain their own frequency and path-mean angle (Fig.~\ref{fig:geometry}). Equation~\eqref{eq:Bdef} then gives
\begin{equation}
 \boxed{\quad d_{ab}=-\bm B_{ab}\cdot\gv,\qquad
                     \gv=K\,\delta\bv .\quad}
 \label{eq:finite}
\end{equation}
The identity is exact for a finite displacement. It requires the path-mean astrometry at the two band edges. Two isolated astrometric epochs generally give a quadrature approximation to that mean. At a single epoch the same result with $\gv=K\dot{\bv}$ holds for the band-averaged \emph{instantaneous} delay drift and the two instantaneous edge separations. A known curved trajectory can be treated by integrating the edge combination against $\dd\bv$ along the curve. A single constant vector then need not describe the full scan.

A general bandpass also follows from Eq.~\eqref{eq:local}. If $w(\nu)$ is normalized to unit integral on $[a,b]$, replace Eq.~\eqref{eq:Bdef} with
\begin{equation}
 \bm B_w=[w\nu\,\overline{\Delta\thv}]_a^b
          -\int_a^b w'(\nu)\nu\,\overline{\Delta\thv}(\nu)\,\dd\nu .
 \label{eq:w}
\end{equation}
For a channel mask, the jumps of $w$ supply additional edge terms. These formulas apply to consistent frequency averaging of channelized group delays. An arrival time from template-matching a frequency-summed, evolving pulse profile can have different weights and must be modeled as its own estimator.

Stack $m$ band and pair measurements as $\bm d$, and form $\Bmat_0$ with rows $\bm B_{ab}\tr$. The hypothesis is $\bm d=-\Bmat_0\gv$. If $\Bmat_0$ has rank $r$, there are $m-r$ linearly independent left-null tests before noise modeling. The product $K\,\delta\bv$ is determined only in the row space, so timing alone does not supply the distance and the angular displacement separately. For a single band with only two independent noncollinear pairs, a two-dimensional displacement fits both exactly. Additional bands or independent pairs can make the relation falsifiable. Cycles made by differencing the same three image times contribute no new measurement. We use a spanning set of image pairs in the experiment below.

\section{What changes with a second plane?}
\label{sec:multi}
\subsection{The endpoint selected by timing}

For two static thin planes, let $\thv_1$ denote the observed image angle and $\thv_2$ the angular coordinate at the second plane. In an isotropic background the multi-plane Fermat construction \cite{Blandford,Petters,Schneider} gives the reduced travel phase in time units
\begin{align}
 F={}&\frac{K_{12}}2|\thv_1-\thv_2|^2
       +\frac{K_{2s}}2|\thv_2-\bv|^2 \nonumber\\
     &+V_1(\thv_1,\nu)+V_2(\thv_2,\nu).
 \label{eq:two}
\end{align}
Here $K_{ij}=(1+z_i)D_iD_j/(cD_{ij})$ and the two screen potentials include their corresponding time scales. For a source variation at fixed frequency,
\begin{align*}
 \delta f_i={}&K_{2s}(\bv-\thv_{2,i})\cdot\delta\bv\\
 &+\sum_{k=1}^2(\nabla_{\theta_k}F|_i)\cdot\delta\thv_{k,i}.
\end{align*}
Both gradients in the sum vanish independently. Subtracting rays gives $\nabla_\beta\Delta f=-K_{2s}\Delta\thv_2$, even if the first crossing is far from the second on the sky. Applying $\partial_\nu(\nu\,\cdot)$ and integrating as in Sec.~\ref{sec:finite} proves the band relation for the terminal crossings. Thus Eq.~\eqref{eq:local} holds with $K_{2s}$ and $\Delta\thv_2$, while ordinary imaging measures $\Delta\thv_1$. The same argument selects the last plane of any longer chain with the terminal term in Eq.~\eqref{eq:two}. This endpoint dependence is a consequence of multi-plane Fermat theory.

One class of two-plane systems is exactly indistinguishable by the unknown-motion test. If $\thv_2=M\thv_1+\bm a$ on the sampled rays with frequency-independent $M$, then $\Bmat_{\mathrm{last}}=\Bmat_0 M\tr$ and $\bm d=-\Bmat_0 M\tr\gv$. The common translation cancels, even if it varies with frequency. A static achromatic foreground with quadratic potential realizes this case. An independent measurement of $K_{2s}\delta\bv$ removes some of this freedom.

\subsection{A recoverable foreground curvature}

Consider a foreground plasma whose phase is quadratic in angle on the region occupied by the rays,
\begin{equation}
 V_1=K_{12}u\left(\tfrac12\thv_1\tr S\thv_1
                         +\bm\ell\cdot\thv_1\right)+v_1(\nu),
 \qquad u=(\nu_\star/\nu)^2 ,
 \label{eq:sheet}
\end{equation}
where $S$ is a dimensionless symmetric matrix and $\nu_\star$ is a specified reference frequency. With the sign convention $V_{\rm plasma}=-uP$, this defines $S=-P_{1,\theta\theta}/K_{12}$. A constant electron column can be added to make the column positive over this region. Only the local quadratic form is needed. The first stationary equation gives
\begin{equation}
 \Delta\thv_2=(I+uS)\Delta\thv_1 .
 \label{eq:map}
\end{equation}
This equation holds for an arbitrary static background screen $V_2$ and finite $S$, on matched branches. It also holds for an affine gradient with a uniform foreground prism because the constant deflection cancels from a pair.

Let $\Bmat_1$ be constructed just like $\Bmat_0$ from measured first-plane angles, with $\nu\,\overline{\Delta\thv_1}$ replaced by $\nu u\,\overline{\Delta\thv_1}$ at each edge. Then
\begin{equation}
 \boxed{\quad \bm d=-\Bmat_0\gv-\Bmat_1\hv,\qquad
              \hv=S\tr\gv,\quad\gv=K_{2s}\delta\bv .\quad}
 \label{eq:extended}
\end{equation}
If $[\Bmat_0\ \Bmat_1]$ has rank four, the four components of $(\gv,\hv)$ are identifiable in the noiseless experiment. One trajectory determines the action of $S$ in one direction. All symmetric matrices obeying $\delta S\tr\gv=0$ remain equivalent. For $S=\epsilon I$ and $\gv\ne0$, the scalar $\epsilon$ follows from $\hv\parallel\gv$, including its sign. The component of $\hv$ orthogonal to $\gv$ can itself test isotropy when the unrestricted fit is identifiable. For two scans with independently identifiable, noncollinear $\gv_1,\gv_2$, define $G=[\gv_1\ \gv_2]$ and $U=[\hv_1\ \hv_2]$. Then $S\tr=UG^{-1}$ and symmetry is a further consistency condition. Converting $S$ into a physical column-density Hessian requires the foreground distance scale and the plasma dispersion coefficient.

For a scalar foreground an even weaker condition suffices locally. Stack the measurements with a fixed positive-definite weight $W$, let $\prj_0$ project orthogonally onto the columns of $W^{1/2}\Bmat_0$, and define $\bm z=\Bmat_1\gv$. At $\epsilon=0$ a small foreground amplitude changes the predicted angular term by $\delta\epsilon\,\bm z$. The displacement fit can absorb a change $\Bmat_0\delta\gv$. In whitened coordinates the best remaining squared residual is
\begin{align*}
 &\min_{\delta\gv}\left\|W^{1/2}
       (\Bmat_0\delta\gv+\delta\epsilon\,\bm z)\right\|^2\\
 &\hspace{0.4cm}=(\delta\epsilon)^2
       \left\|(I-\prj_0)W^{1/2}\bm z\right\|^2.
\end{align*}
The squared length per unit amplitude therefore gives
\begin{equation}
 \mathcal I_\epsilon
   =\big\|(I-\prj_0)W^{1/2}\bm z\big\|^2 .
 \label{eq:info}
\end{equation}
Only the part of the chromatic foreground response outside the single-screen fit carries local information on $\epsilon$. If $\bm z$ lies in the column space, first-order sensitivity vanishes even with more measurements than parameters. The quantity $\mathcal I_\epsilon^{-1/2}$ is a local standard-error estimate for a Gaussian small-noise experiment. We retain errors in the angle measurements and calibrate the finite-sample tests in Sec.~\ref{sec:experiment}.

\subsection{Departure from a quadratic column}

For a general foreground $V_1=-uP_1$, write $\bm a(\thv)=-\nabla P_1(\thv)/K_{12}$. The first stationary equation is exactly $\Delta\thv_2=\Delta\thv_1+u\Delta\bm a$. For each image pair, $\Delta\bm a$ is the angular separation acted on by the average matrix $D\bm a$ along the straight chord joining the two observed positions. A single curvature matrix therefore samples different spatial averages as the frequencies, epochs and pairs change.

In particular, let $\|D\bm a-S\|_{\rm op}\leq\rho$ on all sampled chords and put $m_\nu=\int_0^1\|\Delta\thv_1(s,\nu)\|\dd s$. If $r_{ab}(S)$ is the signed residual of Eq.~\eqref{eq:extended} evaluated at the true $\gv$ and this trial matrix, then
\begin{equation}
 |r_{ab}(S)|\leq \frac{\rho\,|\gv|}{b-a}
           (b u_b m_b+a u_a m_a).
 \label{eq:bound}
\end{equation}
The chord integral gives $\|\Delta\bm a-S\Delta\thv_1\|\leq\rho\|\Delta\thv_1\|$. Taking trajectory means and applying the triangle inequality at the two edges proves the bound. A fitted constant curvature can describe the timing accurately even if it differs from the local value at the foreground center.

\section{What a single source position can determine}
\label{sec:fixed}
\subsection{An inverse on a finite band}

At a fixed $\bv$, suppose the full single-screen phase is affine in $u$,
\begin{equation}
 F(\thv,u)=F_0(\thv)-uP(\thv).
 \label{eq:affine}
\end{equation}
On one stationary branch let $f(u)=F(\thv(u),u)$ and $t(u)=f-2uf'$. The envelope gives $f'=-P$, so the vacuum action and dispersive coefficient sampled on the track are
\begin{equation}
 F_0(\thv(u))=(t+f)/2,\qquad
 P(\thv(u))=(t-f)/(2u).
 \label{eq:sumdiff}
\end{equation}
Each relation also holds for signed pair differences. Write $d=\Delta t$, $f_\Delta=\Delta f$ and $A_\Delta(u)=\Delta[F_0(\thv(u))]$ in this subsection. Since $\nu\partial_\nu u=-2u$, the measured quantity satisfies $d=f_\Delta-2u f_\Delta'$. Dividing by $-2u^{3/2}$ identifies an exact derivative,
\[
 \frac{\dd}{\dd u}\!\left(\frac{f_\Delta}{\sqrt u}\right)
             =-\frac{d(u)}{2u^{3/2}} .
\]
For a finite interval with lower edge $u_a>0$, integration gives
\begin{equation}
 f_\Delta(u)=\sqrt{u}\left[
 \frac{f_\Delta(u_a)}{\sqrt{u_a}}-
 \frac12\int_{u_a}^{u}\frac{d(s)}{s^{3/2}}\,\dd s\right] .
 \label{eq:phaseinverse}
\end{equation}
With a vacuum action supplied independently, the converse is an exact test from positions and delays. Eliminate $f_\Delta$ using Eq.~\eqref{eq:sumdiff}, which gives $ud'-d=2uA_\Delta'-A_\Delta$. Dividing by $u^2$ gives $(d/u)'=2(A_\Delta/u)'+A_\Delta/u^2$. Integrating each term from the same lower edge yields
\begin{align}
 d(u)={}&2A_\Delta(u)
   +u\left[\frac{d(u_a)-2A_\Delta(u_a)}{u_a}
         +\int_{u_a}^{u}\frac{A_\Delta(s)}{s^2}\,\dd s\right].
 \label{eq:vacuumknown}
\end{align}
It needs one delay anchor and the values of a candidate vacuum action along the observed tracks. Neither a Jacobian model nor numerical track derivatives enter this form.

\subsection{A constructive ambiguity}

With freely varying spatial $F_0,P$, the integration constant has a precise physical interpretation. On a compact band bounded away from zero the change
\begin{align}
 f&\longmapsto f+k\sqrt u, \nonumber\\
 F_0(\thv(u))&\longmapsto F_0(\thv(u))+\tfrac k2\sqrt u,\nonumber\\
 P(\thv(u))\longmapsto P(\thv(u))-\frac{k}{2\sqrt u}
 \label{eq:ambiguity}
\end{align}
preserves the group time. It adds a constant $2\pi\nu_\star k$ to the propagation phase in radians. This integration freedom would be merely formal if keeping the same image positions forced $k=0$. The following construction shows when it survives astrometry.

\begin{figure*}[t]
 \centering\includegraphics[width=.91\textwidth]{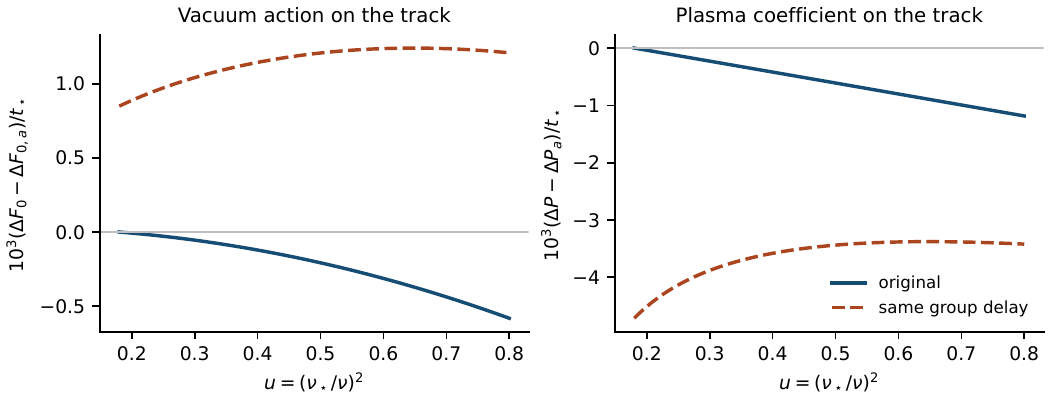}
 \caption{\label{fig:fixed}Two decompositions of the same stationary tracks and pair group delay. We use images 4 and 1 of Sec.~\ref{sec:experiment}, $u_a=0.18$ and $k=0.004\tst$ for the difference in integration constants. Both panels subtract the original value at $u_a$ from both curves. Their changes cancel in $d=\Delta F_0+u\Delta P$ at every displayed point.}
\end{figure*}

Consider a smooth embedded track with nonzero tangent and an interval slightly larger than the measured compact band. In a tubular neighborhood choose a smooth coordinate $s(\thv)$ equal to $u$ on the track. Extend the two perturbations in Eq.~\eqref{eq:ambiguity} as $(k/2)\sqrt{s(\thv)}$ and $-k/[2\sqrt{s(\thv)}]$. Their gradients obey
\begin{equation}
  \nabla\delta F_0=s\,\nabla\delta P .
\end{equation}
A smooth cutoff equal to unity on a neighborhood of the measured segment keeps this equality on the observed rays. Thus every measured position still solves the stationary equation. Taking $|k|$ sufficiently small retains noncriticality. On disjoint observed tracks one can choose the constants independently. Their differences change an inferred pair vacuum action and pair electron column while preserving all measured positions and pair group delays. Figure~\ref{fig:fixed} illustrates the on-track values. This is a local nonuniqueness statement for smooth potentials. Additional images, measured Hessians, a globally prescribed mass model, positivity at a boundary, or global analyticity can constrain the continuation.

If the baseline projected mass density and electron column are strictly positive on the compact supports of the cutoffs, small enough perturbations preserve those inequalities. The construction does not control additional images outside the observed tracks. A calibrated relative phase fixes the constant up to a cycle ambiguity if only phase modulo $2\pi$ is measured and the image phase indices are known. If the data reach $u=0$ on a branch smooth there, regularity of $f'$ also removes the $\sqrt u$ solution. A finite observing band with a smooth spatial model supplies neither boundary condition automatically. An unknown common intrinsic arrival time is invisible to pair timing and leaves additional freedom in individual-branch quantities.

\subsection{Relation to the local Hessian formula}

Let $H=F_{,\theta\theta}$ and $\bm v=\dd\thv/\dd u$. For Eq.~\eqref{eq:affine}, stationarity implies $H\bm v=\nabla P$. Differentiating the envelope gives $-f''=C=\bm v\tr H\bm v$, hence $t''=3C+2uC'$. When $F$ contains a general material term, the corresponding expression is $-f''=C-F_{,uu}$, where the partial derivative holds the angular coordinates fixed. In particular, for $F=F_0-uP-u^2Q+O(u^3)$ near a noncritical vacuum image,
\begin{equation}
 t=t_0+uP_0+u^2\left(3Q_0+\tfrac32\bm v_0\tr H_0\bm v_0\right)
          +O(u^3).
 \label{eq:higher}
\end{equation}
The coefficient can have either sign for a saddle. For a nonnegative cold plasma the local material $Q_0$ is nonnegative before eliminating any hidden ray coordinates. The generalization of Eq.~\eqref{eq:vacuumknown} uses the sampled difference of $F-uF_{,u}$ in place of $A_\Delta$ and is therefore sensitive to $Q$. In contrast, Eqs.~\eqref{eq:local} and \eqref{eq:finite} allow arbitrary local dispersion in $V$.

In a full affine phase with stationary coordinates $(x,y)$, observing only $x$ leaves an additional term. Put $p_y=\partial_yP$ and suppose $H_{yy}$ is invertible. With $H_{\rm eff}=H_{xx}-H_{xy}H_{yy}^{-1}H_{yx}$, eliminating $y'$ from $H(x',y')\tr=\nabla P$ gives
\begin{equation}
 C_{\rm full}=(x')\tr H_{\rm eff}x'+p_y\tr H_{yy}^{-1}p_y .
 \label{eq:schur}
\end{equation}
Indeed, the lower block gives $y'=H_{yy}^{-1}(p_y-H_{yx}x')$. Completing the square in $(x',y')\tr H(x',y')$ gives the displayed result, including the positive sign in front of $p_y\tr H_{yy}^{-1}p_y$. The second term has no fixed sign for indefinite $H_{yy}$. Equation~\eqref{eq:extended} offers a way to constrain a specific hidden-plane effect when source displacement is available.

\section{A four-image experiment}
\label{sec:experiment}
\subsection{Lens, bandpass and independent predictions}

We use a point mass with external shear and an offset Gaussian plasma column on the last plane. Set $\bm x=\thv_2/\theta_E$, $\bm b=\bv/\theta_E$, and $\tst=K_{2s}\theta_E^2$. In these units
\begin{align}
 F_0/\tst={}&\tfrac12|\bm x-\bm b|^2-\ln|\bm x|
                       -\tfrac{\gamma}{2}(x_1^2-x_2^2),\nonumber\\
 P_2/\tst={}&0.08\exp\!\left[-\tfrac12
 \left(\frac{(x_1-0.25)^2}{0.65^2}
       +\frac{(x_2+0.10)^2}{0.45^2}\right)\right] .
 \label{eq:mock}
\end{align}
Here $x_1,x_2$ inside Eq.~\eqref{eq:mock} are Cartesian components in the \emph{same} plane. We set $\gamma=0.18$, $\bm b_0=(0.12,0.04)$, $\delta\bm b=(0.015,-0.010)$ and $\nu_\star=1$ GHz. For a Galactic example, observer-to-plane and observer-to-source distances of $0.30,0.50,1.00$ kpc give $cK_{2s}=1.00$ kpc and $K_{12}/K_{2s}=0.75$. Choosing $\theta_E=10$ mas gives $\tst=241.924\,\mu\mathrm s$, a source displacement of $0.180$ mas, and a peak dispersive coefficient $0.08\tst=19.35\,\mu\mathrm s$ at 1 GHz. Angular positions, times, and $\gv$ below are reported in units of $\theta_E$, $\tst$, and $\tst/\theta_E$, respectively. In those units $\gv=\delta\bm b$. The logarithmic term is normalized to one Einstein radius. With $z_2\simeq0$ its point-mass coefficient satisfies $4GM/c^3=\tst$, giving $M=12.28\,M_\odot$. One physical interpretation is a Galactic compact radio source aligned behind a stellar-mass compact object in an external tidal field, with ionized material on the line of sight. Four images require a suitable source position within the shear caustic. This rare alignment sets the units of a controlled experiment. We do not use it to predict a Galactic event rate.

For the two-plane runs we add Eq.~\eqref{eq:sheet} with $S=\epsilon I$. With $\bm x_o=\thv_1/\theta_E$ and $\eta=K_{12}/K_{2s}$, substitution of $\bm x_o=\bm x/(1+\epsilon u)$ into the phase gives a quadratic addition $\eta\epsilon u|\bm x|^2/[2(1+\epsilon u)]$ to $(F_0-uP_2)/\tst$. The roots of this reduced phase are convenient for continuation. For the timing data we evaluate the \emph{unreduced} group time directly on those roots,
\begin{align}
 t/\tst={}&F_0/\tst+\tfrac{\eta}{2}|\bm x_o-\bm x|^2
   +u\left(P_2/\tst-\tfrac{\eta\epsilon}{2}|\bm x_o|^2\right).
 \label{eq:forward}
\end{align}
A spatially uniform foreground column cancels exactly from all pairs. No angular prediction is used when evaluating Eq.~\eqref{eq:forward}. The angular prediction uses independent quadrature grids for the slow trajectory and only the two frequency endpoints of each band.

We use the four adjacent frequency bands bounded by $(1.00,1.20,1.45,1.70,2.00)$ GHz and the three pairs $(2,1),(3,1),(4,1)$. The four vacuum positions, indexed 1 through 4, are $(1.177,0.087)$, $(-1.019,0.165)$, $(-0.383,0.858)$ and $(-0.295,-0.853)$. We follow these noncritical images and check both equations of the full two-plane model. A separately initialized four-dimensional root solver applied to 180 image configurations agrees with the reduced solution to $3.2\times10^{-14}\theta_E$. A fourth-order finite difference in frequency of $\nu f$ agrees with the group time to $7.6\times10^{-13}\tst$. The smallest absolute eigenvalue of the dimensionless reduced image Hessian over the computed roots exceeds 0.22. Doubling the quadrature order from 12 to 24 for the cases $\epsilon=0,0.06$ changes the drift by less than $2\times10^{-16}\tst$.

\begin{figure*}[t]
 \centering\includegraphics[width=.93\textwidth]{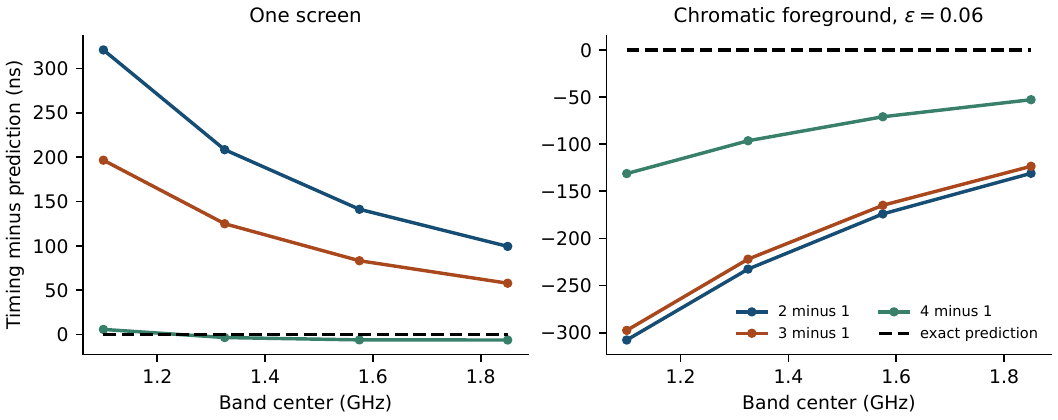}
 \caption{\label{fig:bands}Band-averaged change in pair delay minus the angular prediction, evaluated at the injected source displacement. Left panel: one screen, with the ordinary band-center image separation substituted for Eq.~\eqref{eq:Bdef} (colored curves). The exact band-edge combination gives zero within numerical precision (dashed). Right panel: two screens with $\epsilon=0.06$, using the correct band operator on observed first-plane angles (colored) or on last-plane angles (dashed). The exact residuals of the three pairs coincide at the scale of the plots. Fitting the displacement absorbs part of the mismatch, as quantified in Table~\ref{tab:power}.}
\end{figure*}

In the single-screen run the ordinary band-center image separation gives a largest error of 321 ns. The band-edge rule closes to within $1.2\times10^{-16}\tst$. With $\epsilon=0.06$ the largest error from applying the band-edge rule on the first plane reaches 308 ns. Using the terminal angle, or adding $\epsilon\Bmat_1$ to the observed first-plane prediction, restores agreement within $2\times10^{-16}\tst$ (Fig.~\ref{fig:bands}). The unrestricted four-column fit at this injection gives $(\gv,\hv)=(0.015,-0.010,0.0009,-0.0006)$, with a design condition number of 23.3 in the adopted units. Matrix rank and sensitivity are separate issues: finite errors on the columns affect the distribution of the estimated curvature.

\subsection{Noise, detection power and curvature recovery}
\label{sec:power}

Each simulated per-image, per-epoch, band-averaged time has independent Gaussian error $\sigma_t=24$ ns. Each Cartesian component of each \emph{path-averaged} edge position has independent Gaussian error $\sigma_\theta=4\,\mu$as before pair differencing. Neighboring bands share an astrometric edge and all three pairs share image 1. We propagate both correlations and profile over latent true angles as described in Appendix~\ref{app:stats}. These assumed errors specify the experiment. They do not follow from an instrument or a fit to a known system.

We use two prespecified statistics. $Q_0$ is the minimum residual for a single screen with free two-component displacement. $T=Q_0-\min_\epsilon Q_\epsilon$ targets the signed isotropic quadratic foreground. A gain in $T$ tests this extra degree of freedom within the assumed model family. We generated 2,000 independent null realizations for the 95th-percentile thresholds, obtaining $18.48$ for $Q_0$ and $3.67$ for $T$. The corresponding regular large-sample values for 10 and 1 degrees of freedom are 18.31 and 3.84. We then generated another 1,000 realizations per injected $\epsilon$, including zero. The empirically calibrated thresholds are used throughout Table~\ref{tab:power}.

\begin{table}[t]
 \caption{\label{tab:power}Rejection fractions at nominal 5\% size, and recovered signed isotropic curvature. Each row contains 1,000 independent validation trials. Brackets enclose the 16th and 84th percentiles of fitted values across trials.}
 \begin{ruledtabular}
 \begin{tabular}{cccl}
 $\epsilon$ & $Q_0$ & $T$ & $\widehat\epsilon$ median [16th, 84th]\\
 \hline
 0.00 & .051 & .046 & $-.001\ [-.016,.013]$\\
 0.02 & .090 & .250 & $.020\ [.004,.036]$\\
 0.04 & .248 & .596 & $.040\ [.021,.058]$\\
 0.06 & .466 & .863 & $.061\ [.042,.082]$\\
 0.08 & .643 & .939 & $.080\ [.057,.103]$\\
 0.10 & .756 & .976 & $.100\ [.074,.125]$\\
 \end{tabular}
 \end{ruledtabular}
\end{table}

At $\epsilon=0.06$ the unrestricted single-screen goodness-of-fit test rejects in 466 of 1,000 trials. The targeted test rejects in 863. In the independent null sample the corresponding counts are 51 and 46. The power gain comes from restricting the alternative to the direction specified by Eq.~\eqref{eq:extended}. A failure of the omnibus test to reject an individual realization would therefore have limited force as evidence for a single plane. Conversely, the targeted power does not transfer to other angular profiles without checking their projected response. In the noise-free $\epsilon=0.06$ example, profiling the displacement reduces the single-screen residual to $Q_0=8.55$ in the assumed noise units, below the omnibus threshold, while the true two-plane model has zero residual.

\begin{figure*}[t]
 \centering\includegraphics[width=.93\textwidth]{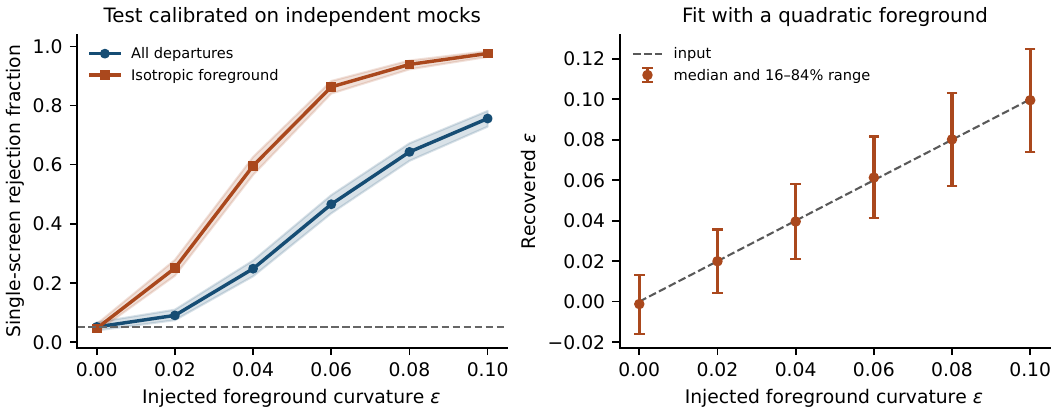}
 \caption{\label{fig:power}Left panel: validation rejection fractions for the two tests. Shading shows 95\% Wilson intervals for binomial sampling conditional on the calibration thresholds. The horizontal dashed line denotes the nominal null rate. Right panel: medians and central 68\% ranges of the recovered $\epsilon$ with unknown displacement. The oblique dashed line denotes the injected value. Recovery ranges describe the distribution of estimators across simulations.}
\end{figure*}

At the null, Eq.~\eqref{eq:info} gives $\sigma_\epsilon=0.0146$, consistent with the central range in Table~\ref{tab:power}. Increasing the path-mean angular uncertainty to $20$ or $100\,\mu$as at the same timing precision increases the local error to 0.0283 or 0.124. For $100$ ns timing and $4\,\mu$as astrometry the local error is 0.0572. The nonlinear sampling distributions in Fig.~\ref{fig:power} broaden with increasing curvature. Extrapolating the null Fisher error to large injections would overstate the precision.

\subsection{A smooth foreground outside the fitted family}

We repeated the entire forward calculation with a positive Gaussian foreground column of angular width $L$. For $w=L/\theta_E$, its phase is $V_1/\tst=-\eta u\epsilon w^2\exp[-|\bm x_o|^2/(2w^2)]$. Every case has the same central curvature $\epsilon=0.060$, while the variation across the image footprint increases as $L$ decreases. A separate four-coordinate solver finds both positions without imposing Eq.~\eqref{eq:map}. Timing is computed from the full two-plane action. Only first-plane positions and pair timing are passed to the quadratic inverse model.

\begin{table}[t]
 \caption{\label{tab:gaussian}Noiseless Gaussian foreground fitted with the isotropic quadratic model and free displacement. The third column is the maximum absolute timing residual at the true displacement and central curvature. The fourth is the maximum after fitting the displacement and scalar curvature.}
 \begin{ruledtabular}
 \begin{tabular}{rccc}
 $L/\theta_E$ & $\widehat\epsilon$ & Central value (ns) & Best fit (ns)\\
 \hline
 8.00 & .05988 & 1.39 & 0.11\\
 4.00 & .05948 & 5.61 & 0.44\\
 2.00 & .05746 & 23.68 & 1.61\\
 1.00 & .04525 & 107.37 & 3.96\\
 0.65 & .02379 & 240.84 & 5.60\\
 \end{tabular}
 \end{ruledtabular}
\end{table}

Table~\ref{tab:gaussian} shows that even an excellent fit can return a curvature substantially different from the central value. At $L=0.65\theta_E$ the fitted value is 0.02379 and the largest post-fit discrepancy is only 5.60 ns. The fitted coefficient describes a projected response across the sampled footprint. Reading it as a pointwise electron-column curvature requires a spatial model or bounds such as Eq.~\eqref{eq:bound}.

Two additional controls test the identified directions. For a chromatic anisotropic quadratic with diagonal entries $(0.060,-0.020)$ and off-diagonal entry $0.018$, a second scan with $\delta\bm b=(0.006,0.017)$ and the original scan give
\begin{equation}
 S\tr=UG^{-1}=
 \begin{pmatrix}0.060&0.018\\ 0.018&-0.020\end{pmatrix}.
\end{equation}
The two design condition numbers are 18.8 and 17.8. The full numerical solutions recover each matrix element within $10^{-10}$ in the absence of noise. For an achromatic quadratic foreground with diagonal entries $(0.08,-0.03)$ and off-diagonal entry $0.02$, the same one-screen fit returns $\widehat{\gv}=(0.016,-0.0094)$ instead of the injected $(0.015,-0.010)$. Its profiled $Q_0$ is below $10^{-22}$. This implements the exactly invisible foreground of Sec.~\ref{sec:multi}.

\subsection{Checks at fixed position}

For a fixed source and a single screen, the four local values of $10^3 C_0/\tst$ are 3.12399, 0.78686, 0.28108 and 1.24029 in image order. The vacuum delay of image 4 relative to image 1 is $0.3749603422\tst$. Fitting its exact group delay at 1,801 equally spaced points in $0.05\leq u\leq0.5$ with a constant plus $u$ biases the intercept upward by $1.67873\times10^{-4}\tst$. Including $u^2$ changes that bias to $-5.38361\times10^{-7}\tst$. Appendix~\ref{app:optical} gives the leading analytic bias. For Fig.~\ref{fig:fixed} the reconstructed phase with its true integration anchor and the direct finite-band timing prediction in Eq.~\eqref{eq:vacuumknown} agree with the forward values within $9\times10^{-16}\tst$. The alternative potential decomposition in that figure produces the same pair time within $6\times10^{-17}\tst$.

\section{Observational interpretation and limits}
\label{sec:discussion}
\subsection{Astrometry, propagation and the noise budget}

A measurement needs the same image branches across frequency and epoch, a matched copy of the same emitted event in each image at each epoch, and relative phase positions in a common angular frame. A common intrinsic frequency sweep cancels within each pair, even when it changes between bursts. An image-dependent scattering tail, frequency-dependent emitting position or blended centroid can bias the test. In particular, an active galactic nucleus used as a phase reference can have an intrinsic chromatic core shift \cite{Lobanov,Porcas}. Common angular translations cancel in separations measured at each edge. Residual differential calibration, source structure and changes in branch identification still require control.

The $4\,\mu$as path-mean precision in Sec.~\ref{sec:power} is demanding. For an $8{,}000$ km projected baseline at 1 GHz, $\lambda/B=7.73$ mas. On a single projected baseline, $\phi=2\pi B\theta_\parallel/\lambda$ and a high signal-to-noise visibility with phase error $\sigma_\phi\simeq1/\rho$ gives $\sigma_{\theta_\parallel}\simeq\lambda/(2\pi B\rho)$. Thus $4\,\mu$as corresponds to a differential path of $0.155$ mm and a thermal phase signal-to-noise ratio of about 310 for that coordinate. A multi-baseline, multi-epoch estimate has its own angular covariance. Ionospheric calibration, source structure and correlated refraction can dominate its thermal error \cite{ReidHonma,Lazio}. In a sample of 58 active galactic nuclei Lazio et al.\ found a typical scattering diameter of 2 mas at 1 GHz \cite{Lazio}. The value is specific to that sample. The apparent width controls blending and correlated flux. The precision of the centroid also depends on signal-to-noise, calibration and refractive wander.

Frequency differencing amplifies angular errors. For one image coordinate with independent, equal path-mean errors $\sigma_\theta$ at two edges, the standard error in Eq.~\eqref{eq:Bdef} before pair differencing is $\sigma_\theta\sqrt{a^2+b^2}/(b-a)$, a factor of 7.81 for $1.00$--$1.20$ GHz. Adjacent bands share an edge. Computing pair delays, positions and Doppler shifts from the same visibilities can add cross-correlations \cite{Cordes,Brisken}. Arrays, scattering monitors and repeated epochs are useful insofar as they constrain this joint covariance. In the four-image geometry, $100\,\mu$as per path-mean coordinate and 24 ns timing give a local curvature error of 0.124, about twice the injected 0.060. A claim of comparable sensitivity for a sky source therefore requires measuring its covariance and checking the projected information in Eq.~\eqref{eq:info}.

We also repeated the complete targeted test at the two coarser angular precisions. For each we recalibrated its 95th-percentile threshold using 2,000 new null trials, then used 1,000 independent trials for each of the null and the injected curvature. The timing error, source trajectory and bands are the same as in Sec.~\ref{sec:power}. Table~\ref{tab:angularpower} shows that a $20\,\mu$as uncertainty lowers the detection fraction to 0.309. At $100\,\mu$as the injected trials cross the threshold in 68 cases, compared with 44 cases in the null sample. At that precision the angular measurements have little power to identify the injected screen in this geometry.

\begin{table}[t]
 \caption{\label{tab:angularpower}Effect of path-mean angular precision in the four-image model. $\sigma_{\epsilon,0}$ is the local error from Eq.~\eqref{eq:info}. $P_0$ and $P_{.06}$ are simulated targeted rejection fractions at true $\epsilon=0$ and $0.06$. The last column gives the 16th and 84th percentiles of fitted curvatures at the injection. The first row is reproduced from Sec.~\ref{sec:power}. Each fraction uses 1,000 validation trials.}
 \begin{ruledtabular}
 \begin{tabular}{rcccc}
 $\sigma_\theta$ ($\mu$as) & $\sigma_{\epsilon,0}$ & $P_0$ & $P_{.06}$ & $[q_{16},q_{84}]$\\
 \hline
 4   & .0146 & .046 & .863 & $[.042,.082]$\\
 20  & .0283 & .042 & .309 & $[.022,.098]$\\
 100 & .124  & .044 & .068 & $[-.106,.258]$\\
 \end{tabular}
 \end{ruledtabular}
\end{table}

\subsection{Galactic pulsars and pure plasma lenses}
\label{sec:pulsars}

A compact mass is optional in Eq.~\eqref{eq:screen}. A bright pulsar whose stable scattering paths can be associated across frequency is a candidate for a first test. Plasma refraction has resolved emitting regions of PSR B1957+20 during eclipse \cite{Main}. That source-plane resolving power concerns the size of the emission region. The measurement proposed here requires the sky directions of the arriving rays. PSR B0834+06 provides an example of interferometric mapping of scattering structure at 327 MHz, at roughly $100\,\mu$as angular resolution \cite{Brisken}. The evolving scintillation of PSR B1508+55 has been modeled with two screens \cite{Sprenger}. Its evolution makes the choice of a time window in which the pattern is effectively frozen particularly important.

A gravity-free calculation illustrates what could be measured. Take a one-dimensional, overdense Gaussian plasma lens \cite{Clegg,CordesFRB}. Let $\theta_p$ denote its angular width, $x=\theta/\theta_p$, $y=\beta/\theta_p$, $T_p=K\theta_p^2$ and $u=(1\,{\rm GHz}/\nu)^2$. With $A>0$ the normalized phase, lens equation and group delay are
\begin{align}
 F/T_p&=(x-y)^2/2-Au e^{-x^2/2},\nonumber\\
 y&=x(1+Au e^{-x^2/2}),\nonumber\\
 t/T_p&=(x-y)^2/2+Au e^{-x^2/2}.
 \label{eq:plasmaonly}
\end{align}
Set $A=10$, $\theta_p=1$ mas and $cK=1$ kpc, so $T_p=2.419\,\mu{\rm s}$. In a Galactic realization with $D_L=0.50$ kpc and $D_S=1.00$ kpc, the Gaussian width is 0.50 AU. Its peak plasma delay at 1 GHz is $24.19\,\mu{\rm s}$, equivalent to an excess dispersion measure of $5.83\times10^{-3}\,{\rm pc\,cm^{-3}}$ through the usual cold-plasma relation \cite{CordesFRB}. The source moves from $y=3.55$ to $3.61$. For an effective transverse speed of $50\,{\rm km\,s^{-1}}$ projected onto the lens plane, the 0.030 AU scan lasts about 25 h. The band edges are 1.00, 1.12, 1.25 and 1.35 GHz. To check branch continuity, the two positive turning points solve $1+Au(1-x^2)e^{-x^2/2}=0$. At a turning point the derivative of its mapped source position with respect to $Au$ is $x e^{-x^2/2}>0$. Over the whole band the largest lower fold lies at $y\simeq3.33$ and the smallest upper fold at $y\simeq4.41$. The entire source scan lies between them. We bracket all three images on their separate monotone intervals and label them by increasing $x$.

For images 2 and 1 in the lowest band, integrating their group times gives $+76.56$ ns. Their path-mean separation falls from 2.171 to 1.882 mas across the band. Consequently $b\,\overline{\Delta\theta}(b)<a\,\overline{\Delta\theta}(a)$ even though the separation stays positive. Using the two edges gives the same positive drift, whereas the band-center separation of 2.028 mas gives $-294.32$ ns. Over two independent image pairs and three bands, the angular and timing calculations agree to $10^{-8}$ ns or better in this noiseless example. A doubled quadrature order changes the direct results by less than $10^{-8}$ ns. These tolerances concern numerical evaluation only. At 1 GHz the phase scale $2\pi\nu T_p$ exceeds $1.5\times10^4$. With a one-dimensional design, one component of source displacement is fitted across six rows. The other component is unmeasurable. Table~\ref{tab:plasmaband} gives all six drifts. The code and the unrounded edge and timing measurements are provided in the Supplemental Material \cite{Supplement}.

\begin{table}[t]
 \caption{\label{tab:plasmaband}Band-edge astrometry and delay drifts for the gravity-free lens in Eq.~\eqref{eq:plasmaonly}. The two angular separations are path means in mas. The last two columns are in ns. $d_{ab}$ is the drift obtained by integrating the group delays. The band-edge law reproduces it to the displayed precision in every row. $d_{\mathrm{mid}}$ instead uses the path-mean angle at the band center.}
 \begin{ruledtabular}
 \setlength{\tabcolsep}{3pt}
 \begin{tabular}{ccrrr}
 Band (GHz) & Pair & $(\overline{\Delta\theta}_a,\overline{\Delta\theta}_b)$ & $d_{ab}$ & $d_{\rm mid}$\\
 \hline
 1.00--1.12 & $(2,1)$ & $(2.171,1.882)$ & $+76.56$ & $-294.32$ \\
 1.00--1.12 & $(3,1)$ & $(3.161,3.090)$ & $-362.21$ & $-454.08$ \\
 1.12--1.25 & $(2,1)$ & $(1.882,1.551)$ & $+189.70$ & $-249.69$ \\
 1.12--1.25 & $(3,1)$ & $(3.090,2.985)$ & $-303.08$ & $-441.46$ \\
 1.25--1.35 & $(2,1)$ & $(1.551,1.264)$ & $+336.79$ & $-204.98$ \\
 1.25--1.35 & $(3,1)$ & $(2.985,2.881)$ & $-229.17$ & $-426.26$ \\
 \end{tabular}
 \end{ruledtabular}
\end{table}

For pair $(2,1)$ the positive separation falls sufficiently rapidly with frequency to reverse the band-center prediction in all three bands. For pair $(3,1)$ its decrease is shallower, so the band-center approximation preserves the sign but overestimates the magnitude of each drift. This comparison isolates the error caused by using the band center in a lens with a single dispersion screen.

The largest path-mean pair separation at the tabulated edges is 3.16 mas, below the 7.73 mas fringe spacing of the single 1 GHz baseline above. The example therefore requires separating paths in delay or interference phase, or obtaining sufficient angular resolution by other means. Gated voltages or matched pulse copies can supply timing, and multi-station phase information can supply relative positions. Inferring angles from the same delays under an assumed screen distance and a one-screen mapping would impose part of the relation being tested. Multi-station data can supply independent geometric information even if both estimators share noise. For an eclipsing binary the source trajectory may curve appreciably; short orbital segments or a known trajectory are needed. In an extreme scattering event of a continuum source, a flux spectrum constrains refraction \cite{PenKing,Tuntsov,Simard}, but supplies no pair delay by itself. A resolved pulse, a repeated temporal feature, or an independently calibrated interference delay would supply the missing observable.

\subsection{Strongly lensed repeating fast radio bursts}

A galaxy-scale lens offers a different regime. Macroscopic images may be separated by an arcsecond, and matched copies of successive bursts can track their delay over time \cite{DaiLu,Zitrin,Wucknitz}. Consider a distance factor $D_{\Delta t}=cK=1$ Gpc and a relative angular source displacement of $0.1\,\mu$as. For one arcsecond of projected separation, the absolute achromatic drift is $0.242$ s. To isolate the dispersive scale, compare bands $p,q$ for the same pair and trajectory. Equation~\eqref{eq:finite} gives
\begin{equation}
 d_p-d_q=-\frac{D_{\Delta t}}{c}\,
       \delta\bv\cdot(\bm B_p-\bm B_q).
 \label{eq:frb}
\end{equation}
The common achromatic separation cancels. A $1$ mas difference of the two band-edge combinations projected along the assumed displacement then gives $0.242$ ms. Both estimates scale linearly with distance factor, projected angle and displacement. The latter is a measurable relative shift or a fitted nuisance parameter. For perspective, $0.1\,\mu$as at an angular-diameter distance of 1 Gpc corresponds to 100 AU of transverse displacement. These are illustrative scales. A particular repeater must supply the astrometry and motion over the available observing interval.

With two macroscopic images, a single band difference gives only one scalar constraint. Several bands with angular leverage, or additional images, are needed for a test with unknown two-dimensional motion. A timing echo from an angularly unresolved compact lens can probe lens masses \cite{Munoz} but by itself supplies no independent image separation for Eq.~\eqref{eq:finite}. Burst association, frequency-dependent emitting positions, differential scattering in the host or lens galaxy \cite{CordesFRB}, and frequency-dependent selection of bright subimages should be checked before interpreting a residual.

\subsection{Screen evolution and caustics}

Slow changes in the lens provide a further ambiguity. If $V$ depends explicitly on normalized epoch $s$, the local drift acquires $\partial_\nu[\,\nu\,\Delta V_{,s}\,]$, with $V_{,s}$ evaluated on the rays, in addition to the displacement term. For example, an evolving achromatic tilt $-K\thv\cdot[\bv(s)-\bv_0]$ at fixed $\bv_0$ reproduces a translated source in both the image equations and the pair delays. Observer acceleration, independent screen velocities and secular distance changes need their own terms. A failed test rejects the joint hypothesis of static propagation and a common achromatic source path. The achromatic foreground of Sec.~\ref{sec:multi} gives an explicit two-plane geometry that passes the test.

Near a caustic, separate stationary branches can merge within the observing band. The detector then measures a coherent superposition with frequency-dependent amplitudes. Uniform wave asymptotics near folds, diffraction integrals, and interference methods for lens dimensionality describe aspects of this regime \cite{Grillo,Feldbrugge,Jow}. An individual group delay in Eq.~\eqref{eq:local} can be assigned only while the branch remains identifiable throughout the integration domain. Modeling a merger requires a specified wave observable and band response.

\section{Conclusions}

For a frozen thin screen a finite displacement changes an image-pair group delay by the edge-weighted angular separation dotted into the displacement, Eq.~\eqref{eq:finite}. The formula uses quantities that can in principle be measured without fitting the gravitational mass or electron column. The gravity-free example makes the distinction operational: one image pair has a positive band drift even though its band-center separation predicts a negative drift.

With a second screen, timing follows the last ray crossing. Fitting a common unknown motion leaves a measurable projection of quadratic foreground curvature, Eq.~\eqref{eq:extended}. Two independent, identifiable scans can recover the full symmetric tensor. A frequency-independent affine remapping remains invisible, and a varying foreground can fit a constant-curvature law with small timing residuals while biasing its inferred central curvature. A fixed source has a different ambiguity: smooth potentials can preserve the tracks and group times while changing their vacuum and material decomposition, Eq.~\eqref{eq:ambiguity}.

A first observational test can use identified scattering paths of a bright pulsar or resolved macroscopic images of a repeating FRB. In each case the decisive measurements are angles at both band edges, repeated pair timing, and their covariance. The projected response after fitting the displacement determines whether a given dataset can distinguish the two geometries.

\begin{acknowledgments}
The author used OpenAI ChatGPT (GPT-6) to assist with algebraic steps, simulation and plotting code, and drafting of the text.
The author directed this use, independently verified all derivations, code and numerical results, and takes full responsibility for the content.
\end{acknowledgments}

\section*{Data availability}

All numerical data in this article are synthetic. The Supplemental Material \cite{Supplement} contains the scripts, pinned software versions, random seed, simulated trials, numerical cross-checks and data underlying the tables and figures. The data archive is supplied separately as \texttt{\detokenize{Guo_Herts_PRD_v4_data.zip}}. Its \texttt{README.TXT} gives the commands and identifies the input and output of each calculation.

\appendix
\section{Optical invariants and the fixed-source expansion}
\label{app:optical}

Consider a static isotropic metric $\dd s^2=-A c^2\dd t^2+B\,\dd\bm x^2$, with positive $A,B$ and a specified reference clock. In a cold, isotropic, unmagnetized test plasma with local plasma angular frequency $\omega_p$, the Hamiltonian for an eikonal of conserved angular frequency $\omega$ is \cite{Perlick,PlasmaReview}
\begin{equation}
  2\mathcal H=-\omega^2/(Ac^2)+|\bm k|^2/B+\omega_p^2/c^2=0 .
\end{equation}
Writing $N=\sqrt{B/A}$ and $Y=A\omega_p^2$ gives, for $\omega^2>Y$,
\begin{align}
 n_{\rm ph}&=N\sqrt{1-Y/\omega^2},\nonumber\\
 n_g&=\partial_\omega(\omega n_{\rm ph})
       =N/\sqrt{1-Y/\omega^2}.
\end{align}
Spatial rays make the phase $\omega c^{-1}\int n_{\rm ph}\dd l$ stationary. The group time is its frequency derivative along a stationary branch. A smooth $\Omega>0$, equal to one near the source, observer, and calibration clocks, gives the transformation
\begin{equation}
 (A,B,\omega_p^2)\longmapsto
  (\Omega^2A,\Omega^2B,\Omega^{-2}\omega_p^2).
 \label{eq:conformal}
\end{equation}
Both indices are invariant pointwise. The Hamiltonian rescales by $\Omega^{-2}$ and has the same characteristics on its zero surface up to reparametrization. Endpoint angles and proper differential times are preserved. Thus the unrestricted propagation problem cannot separate the metric lapse from the spatial scale and free plasma density. Vacuum propagation recovers the usual conformal freedom \cite{Perlick}. Interior clocks, independent density information or field equations with specified matter can restrict the family. Models of gravitational slip that impose dynamics have different input information \cite{Slip}.

On a prescribed static reference-ray family the weak-dispersion coefficients are $P=(2c\omega_\star^2)^{-1}\int NY\,\dd l$ and $Q_{\rm mat}=(8c\omega_\star^4)^{-1}\int NY^2\,\dd l$. The group integrand has a different series from the phase integrand. In cosmological reductions these local expressions need the local frequency and redshift conversion to observer time. In the weak-field vacuum limit the leading index depends on the sum of the two metric potentials. Adding an achromatic term $\chi(\thv)$ to the projected vacuum action shifts the lens equation by $K^{-1}\nabla\chi$. Keeping the mapping on an open connected angular domain forces $\chi$ to be constant there. At finitely many images distinct smooth bumps that are locally constant can change the potential values while preserving the measured positions and local Jacobians. Physical density priors and constraints on other images can restrict such bumps.

There is a separate projected mass-sheet freedom. For a single source plane, $\psi\to\lambda\psi+(1-\lambda)|\thv|^2/2$, $\bv\to\lambda\bv$ scales the dimensionless Fermat differences \cite{Falco,SchneiderSluse}. If in addition $K\to K/\lambda$ and the material potential in time units is fixed, the full screen changes only by a source-dependent common term. The product $\gv$ in Eq.~\eqref{eq:finite} stays invariant. Constraining this product alone therefore cannot break the mass-sheet and distance degeneracy.

For Eq.~\eqref{eq:higher} put $\bm p=\nabla P(\thv_0)$, $\bm v_0=H_0^{-1}\bm p$. Taylor expansion of the stationary branch gives
\begin{equation}
 f=t_0-uP_0-u^2\left(Q_0+\tfrac12\bm p\tr H_0^{-1}\bm p\right)+O(u^3).
\end{equation}
Applying $1-2u\partial_u$ supplies the factor of three in the group time. For a pair write $d=d_0+d_1u+d_2u^2+O(u^3)$, with $d_2=3\Delta Q_0+3\Delta C_0/2$. If one fits just the first two columns $X=(\bm1,\bm u)$, the leading coefficient bias is
\begin{equation}
 \delta\widehat{\bm d}_{01}
 =d_2(X\tr W X)^{-1}X\tr W\bm u^{\,2},
\end{equation}
where the square in $\bm u^{\,2}$ acts elementwise. For uniform continuum weight in $u\in[l,h]$, the intercept bias is $-d_2(l^2+4lh+h^2)/6$. Real channel weights or covariance must be used for an observed band.

\section{Likelihood and finite sampling}
\label{app:stats}

For $n$ adjacent bands define the difference matrices $L_p$, $p=0,1$, acting on the $n+1$ edge coordinates. For edges $\nu_j$, $u_j=(\nu_\star/\nu_j)^2$, their two nonzero entries per row are
\begin{align}
 (L_p)_{j,j}&=-\nu_j u_j^p/(\nu_{j+1}-\nu_j),\nonumber\\
 (L_p)_{j,j+1}&=\nu_{j+1} u_{j+1}^p/(\nu_{j+1}-\nu_j).
\end{align}
$E$ forms the three reference-image differences from four image values, with rows $\bm e_i\tr-\bm e_1\tr$ for $i=2,3,4$. In band-major order the covariance of the elementary pair differences contains $J=EE\tr$ with diagonal 2 and off-diagonal 1. Let $L_\epsilon=L_0+\epsilon L_1$. In consistent angle and time units the residual for a trial $\gv,\epsilon$ and its sampling covariance at the true parameters are
\begin{align}
 \bm r&=\widehat{\bm d}
             +(\widehat{\Bmat}_0+\epsilon\widehat{\Bmat}_1)\gv,\nonumber\\
 \Sigma&=\left[2\sigma_t^2 I_n+
   \sigma_\theta^2|\gv|^2L_\epsilon L_\epsilon\tr\right]\otimes J .
 \label{eq:cov}
\end{align}
The leading factor 2 comes from two independent timing epochs. The angle error is on the already averaged astrometry. It carries no such factor. Both $L_0$ and $L_1$ use the same noisy edge positions, so their errors are correlated. When timing and astrometry are correlated, their cross-covariance also enters the propagated residual.

To see why $\bm r\tr\Sigma^{-1}\bm r$ can be minimized over the motion in this experiment, collect the unknown true edge coordinates into $\bm x$, with covariance $C_x$ for $\widehat{\bm x}$. For given $\gv,\epsilon$ let $A$ be the linear map from those edge coordinates to the angular term in the residual. Minimizing the joint Gaussian quadratic
\begin{align}
 &(\widehat{\bm d}+A\bm x)\tr C_d^{-1}(\widehat{\bm d}+A\bm x)
   \nonumber\\[-2pt]
 &\quad+(\widehat{\bm x}-\bm x)\tr C_x^{-1}
                         (\widehat{\bm x}-\bm x)
\end{align}
over $\bm x$ gives $\bm r\tr(C_d+A C_xA\tr)^{-1}\bm r$. This is a profile over fixed, unknown true positions. The normalizations of the two elementary Gaussian measurements are constant under this profiling. An analysis marginalizing the positions under a population prior has a different likelihood. We minimize the profile quadratic at $\epsilon=0$ for $Q_0$ and over signed $\epsilon$ for the targeted statistic. For Eq.~\eqref{eq:info} we use $W=\Sigma(\gv,0)^{-1}$ at the true null parameters. A broad search over many foreground profiles would require its own null calibration.

The finite-path integral in Eq.~\eqref{eq:Bdef} can be obtained from intermediate astrometric epochs or replaced by a controlled approximation. For uniform epoch spacing $h_s$ in normalized $s$, a composite trapezoid with $M_\nu=\sup_s\|\partial_s^2\Delta\thv(s,\nu)\|$ gives
\begin{equation}
 \|\delta\bm B_{ab}\|\le
       \frac{h_s^2}{12(b-a)}(bM_b+aM_a).
\end{equation}
This follows by applying the ordinary trapezoid remainder at each edge. Similarly, for channel spacing $h_\nu$ the frequency-averaged timing error is at most $h_\nu^2\sup_\nu|\partial_\nu^2[\Delta t_1-\Delta t_0]|/12$, where the subscripts here indicate the two epochs. The bound requires curvature control between samples. In the one-screen benchmark, using only the two temporal endpoints gives a maximal drift-prediction quadrature error of 1.52 ns. Four equal temporal subintervals reduce it to 0.095 ns. Using only the two frequency edges to approximate the \emph{timing} integral gives 4.59 ns, falling to 0.291 ns with four subintervals. These numbers refer to distinct integrations. The astrometric band-edge identity itself introduces no frequency-quadrature error for a top-hat band.

\end{document}